\documentclass[11pt]{article}
\usepackage{moriond,epsfig}

\def\be{\begin{equation}}
\def\ee{\end{equation}}
\def\bea{\begin{eqnarray}}
\def\eea{\end{eqnarray}}

\begin{document}
\vspace*{4cm}
\title{PRECISION FLAVOR PHYSICS FROM THE LATTICE}

\author{C.~Hoelbling}

\address{Bergische Universit\"at Wuppertal, Gaussstr.\,20, D-42119
Wuppertal, Germany.}

\author{for the\\Budapest-Marseille-Wuppertal Collaboration}

\address{}

\maketitle\abstracts{We present results for the light quark masses and
the neutral kaon mixing parameter $B_K$ from lattice QCD. Our data set
includes lighter than physical light quark masses and 5 lattice
spacings so that chiral extrapolation is not necessary and cutoff
effects are fully under control. We obtain fully nonperturbative
predictions for $m_{ud}=(m_u+m_d)/2$, $m_s$ and $B_K$ in the RI scheme
with $M_\pi$, $M_K$ and $M_\Omega$ as the only input quantities. Using
perturbative 4-loop respectively 2-loop running and dispersive input from
$\eta\rightarrow 3\pi$, we obtain
$m_u^\mathrm{\overline{MS}}(2~\mathrm{GeV})=2.17(4)(10)~\mathrm{MeV}$,
$m_d^\mathrm{\overline{MS}}(2~\mathrm{GeV})=4.79(7)(12)~\mathrm{MeV}$,
$m_s^\mathrm{\overline{MS}}(2~\mathrm{GeV})=95.5(1.1)(1.5)~\mathrm{MeV}$
and $\hat{B}_K=0.773(8)(8)$ where the first error is statistical and the second systematic.}

\section{Introduction}

Lattice QCD is a tool to perform ab-initio calculations of QCD in the
nonperturbative regime. One can stochastically perform the functional
integral over gauge and fermion fields on a lattice regulated theory
in finite volume. In order to obtain QCD predictions, it is then
necessary to (a)~remove the cutoff (b)~extrapolate to infinite volume
(c)~tune the bare parameters of the theory or interpolate/extrapolate
to these bare parameter values such that a predefined set of
experimentally accessible, dimensionless quantities (e.g. hadron mass
rations) assume their physical value. If these requirements are met,
one can use lattice QCD to obtain ab-initio QCD predictions of
quantities not used as input in step (c) with a statistical error that
arises from the stochastic integration.

Generally speaking, the challenge for lattice QCD is to simultaneously
fulfill all the requirements (a)-(c) with control over systematic
errors arising from each step and to minimize the total (statistical
plus systematic) uncertainty on a target quantity with given computer
resources. It has been shown recently, that relatively straightforward
quantities such as the ground state light hadron spectrum can be
reproduced with a few percent accuracy.\cite{Durr:2008zz}
Here we present the results of a determination of the light and
strange quark masses as well as for the neutral kaon mixing parameter
$B_K$ with controlled errors on the percent level. For the full
technical details, we refer the reader to the original publications
\cite{Durr:2010vn,Durr:2010aw,Durr:2011ap}.

\section{Quark masses}

Light quark masses are fundamental parameters of the standard model
Lagrangian that are inaccessible by direct experiment. In order to
compute them, we compute some light hadron masses in lattice QCD with
a number of bare input light quark masses. We then interpolate the
bare quark masses at one value of the bare coupling $g$ to the point
where the measured light hadron masses take on their physical value
and renormalize them. We compute the renormalization constant
nonperturbatively in the RI-MOM scheme \cite{Durr:2010aw,Martinelli:1994ty}.

\begin{figure*}
\centerline{\includegraphics*[width=5.5cm]{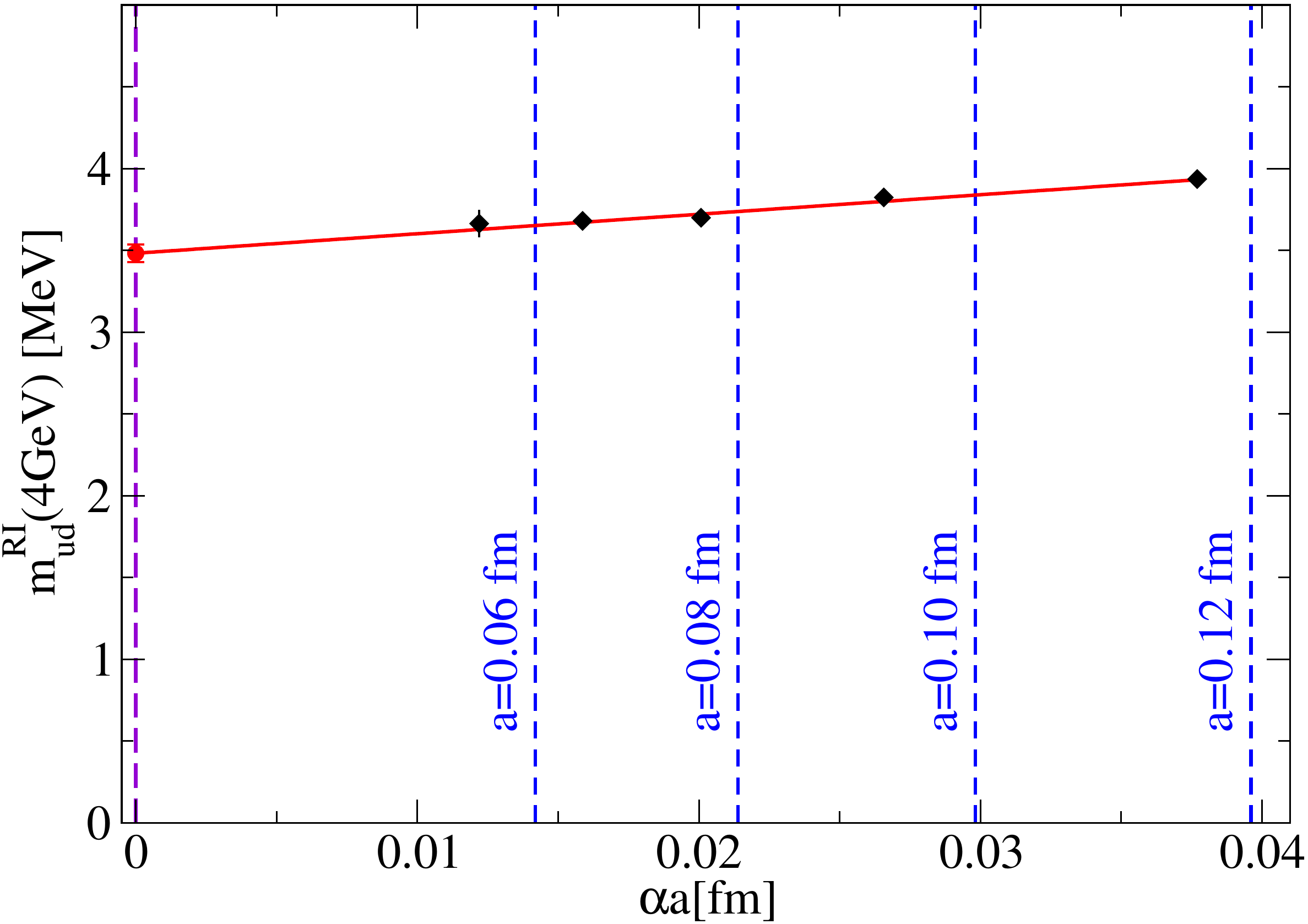}\hspace*{4mm}
\includegraphics*[width=5.7cm]{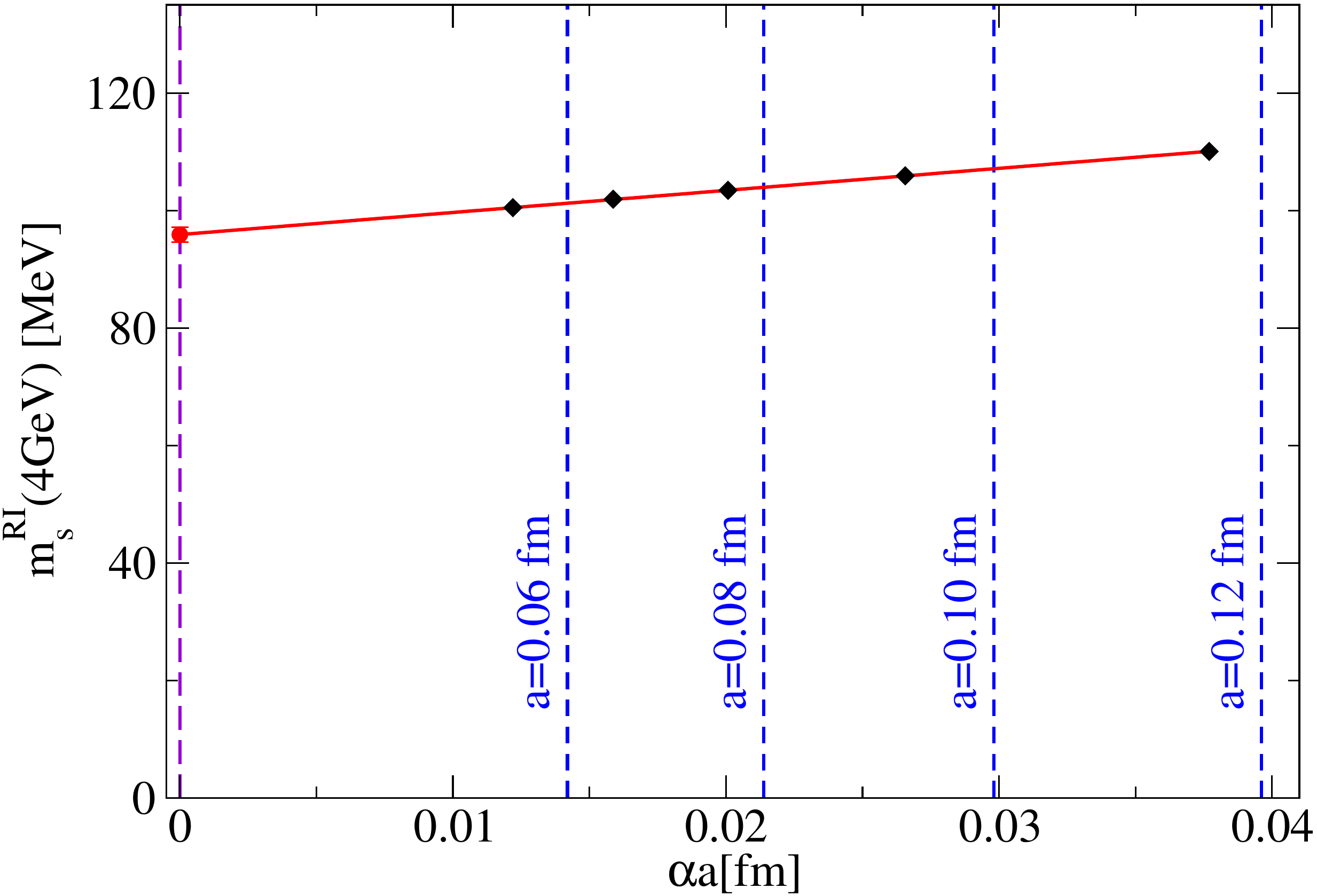}\hspace*{4mm}
\includegraphics*[width=5.5cm]{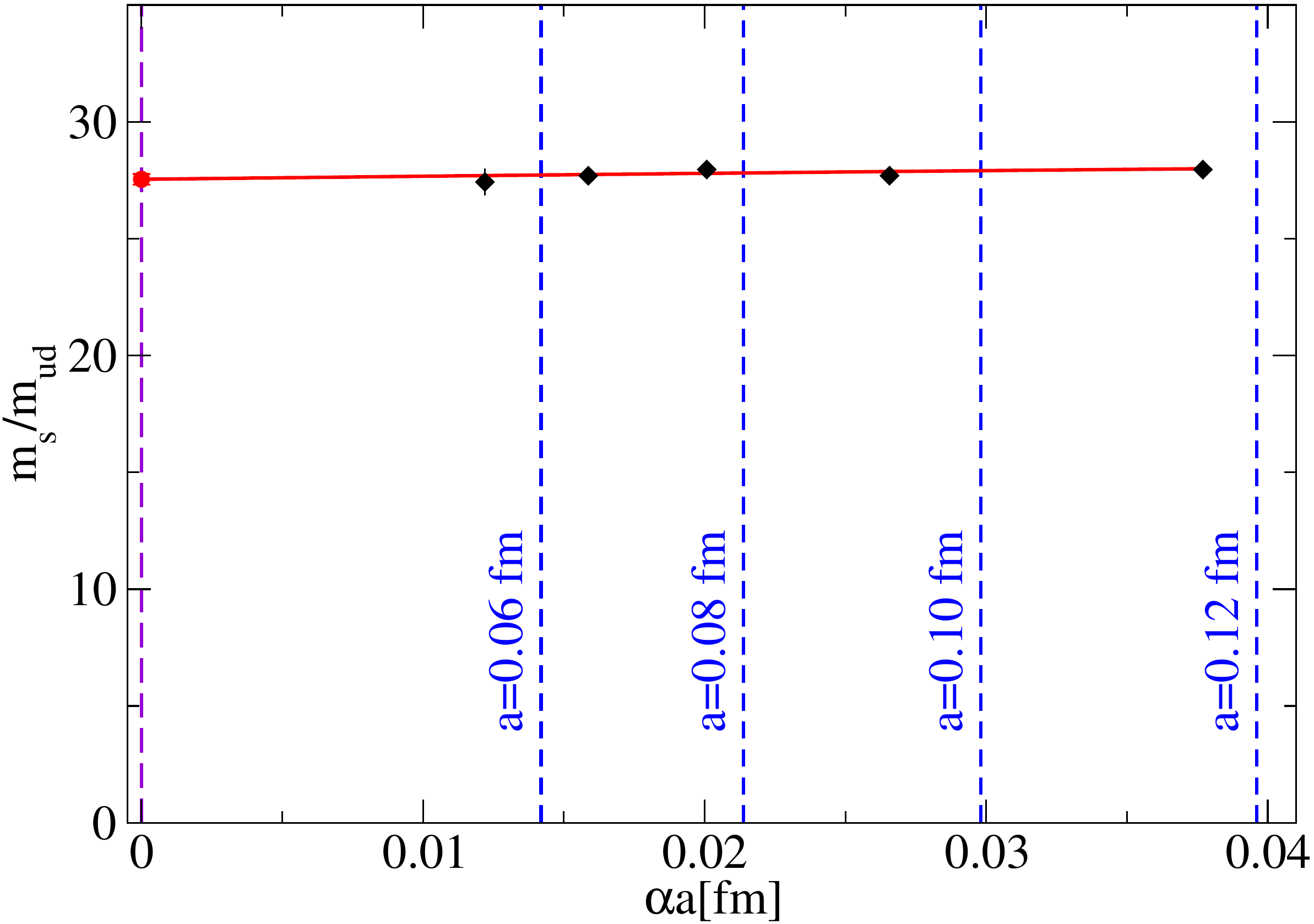}\hspace*{4mm}}
  \caption{
  \label{fig:mqa} Continuum extrapolation of the average up/down quark mass,
   of the strange quark mass and of the ratio of the two.}
\end{figure*}

For technical reasons we work in the isospin limit and correct for the
small isospin breaking at a later stage. We therefore take as input
for finding the physical point the ratios of isospin averaged hadron
masses $M_\pi/M_\Omega$ and $M_K/M_\Omega$. The lattice cutoff or,
equivalently, the lattice spacing $a$ at the given bare coupling $g$
itself is determined by comparing the dimensionless mass of the
$\Omega$ baryon as measured on the lattice to the physical mass
$M_\Omega$. We interpolate to the physical point with various
functional forms.\cite{Durr:2010aw} The resulting spread enters (as a
subdominant part) into our systematic error. Repeating this procedure
at various different values of the bare coupling (in our case we use
5), we get the renormalized quark mass as a function of the lattice
spacing $a$ that we can extrapolate to $a=0$ (see fig.~\ref{fig:mqa}).

For our lattice action, the leading term in the continuum
extrapolation is formally of $O(\alpha_s a)$. There is however strong
evidence that this term is numerically subdominant (for lattice
spacings we work at) to the term of
$O(a^2)$.\cite{Hoffmann:2007nm} We therefore use both
forms in our analysis and include the resulting spread into our
systematic error.

\begin{figure*}
\centerline{\includegraphics*[width=12cm]{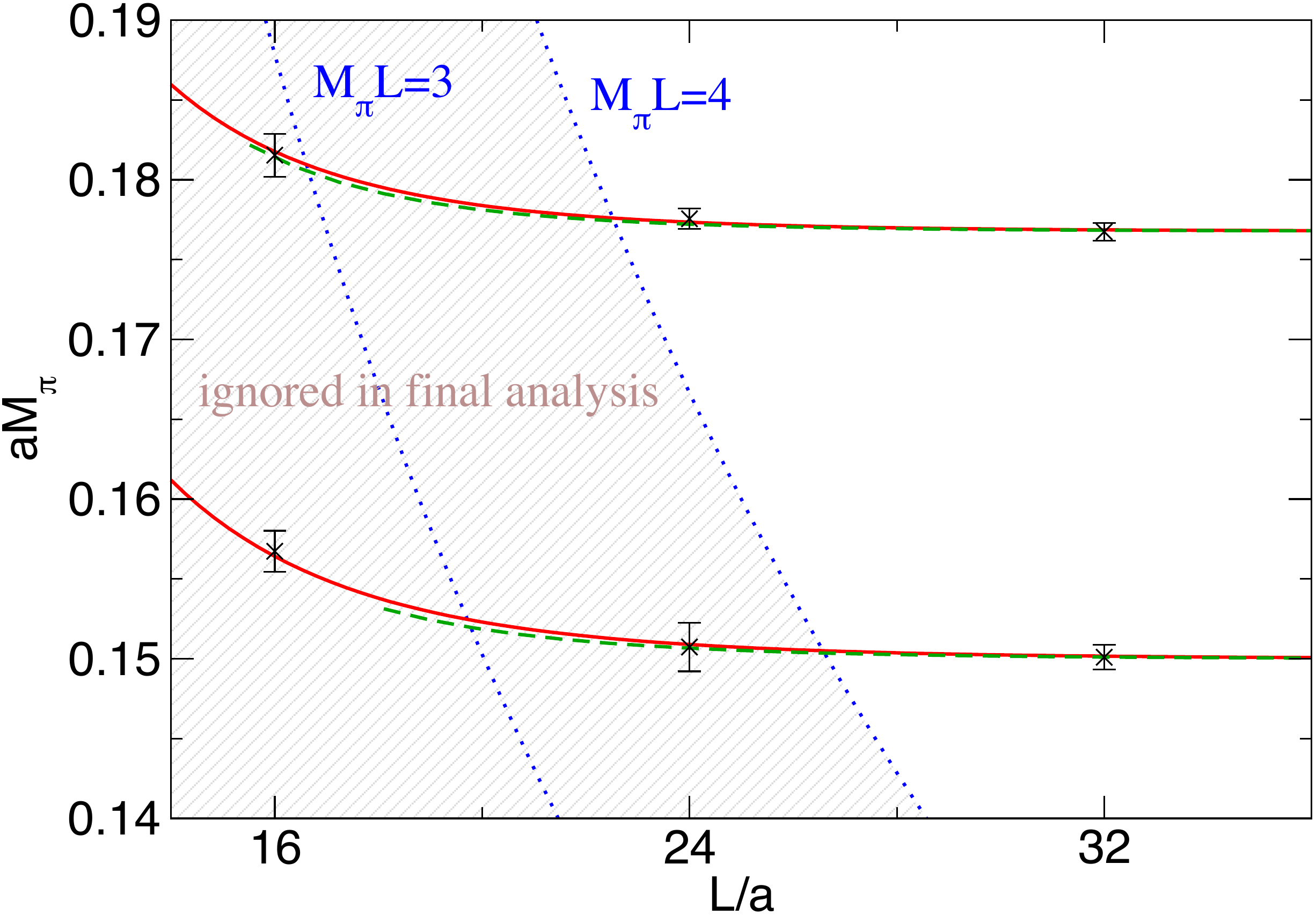}}
\caption{
 \label{fig:fv} Finite volume effects in our data compared to analytic
  predictions. The dashed line corresponds
  to an approximate 3-loop prediction while the full line is an
  asymptotic approximation with one free coefficient.}
\end{figure*}

Finite volume corrections to stable particle masses are generically
exponentially suppressed in $M_\pi L$ and can be corrected for
systematically\cite{Colangelo:2005gd} (see fig.~\ref{fig:fv}). We
included these corrections and found them generically to be at the
permil level.

In order to obtain individual masses for the up and the down quark, we
use the double ratio \be Q^2=\frac{m_s^2-m_{ud}^2}{m_d^2-m_u^2} \ee to
convert our precise result for $m_s/m_{ud}=27.53(20)(8)$ into an
estimate of $m_u/m_d$. In principle, $Q$ can be determined
experimentally from $\eta\rightarrow 3\pi$ decays via dispersion
relations. Due to the imperfect experimental data, there is some
amount of modeling involved and we use a conservative estimate
$Q=22.3(8)$ from a recent review.\cite{Leutwyler:2009jg}

\begin{figure*}
\centerline{\includegraphics*[width=12cm]{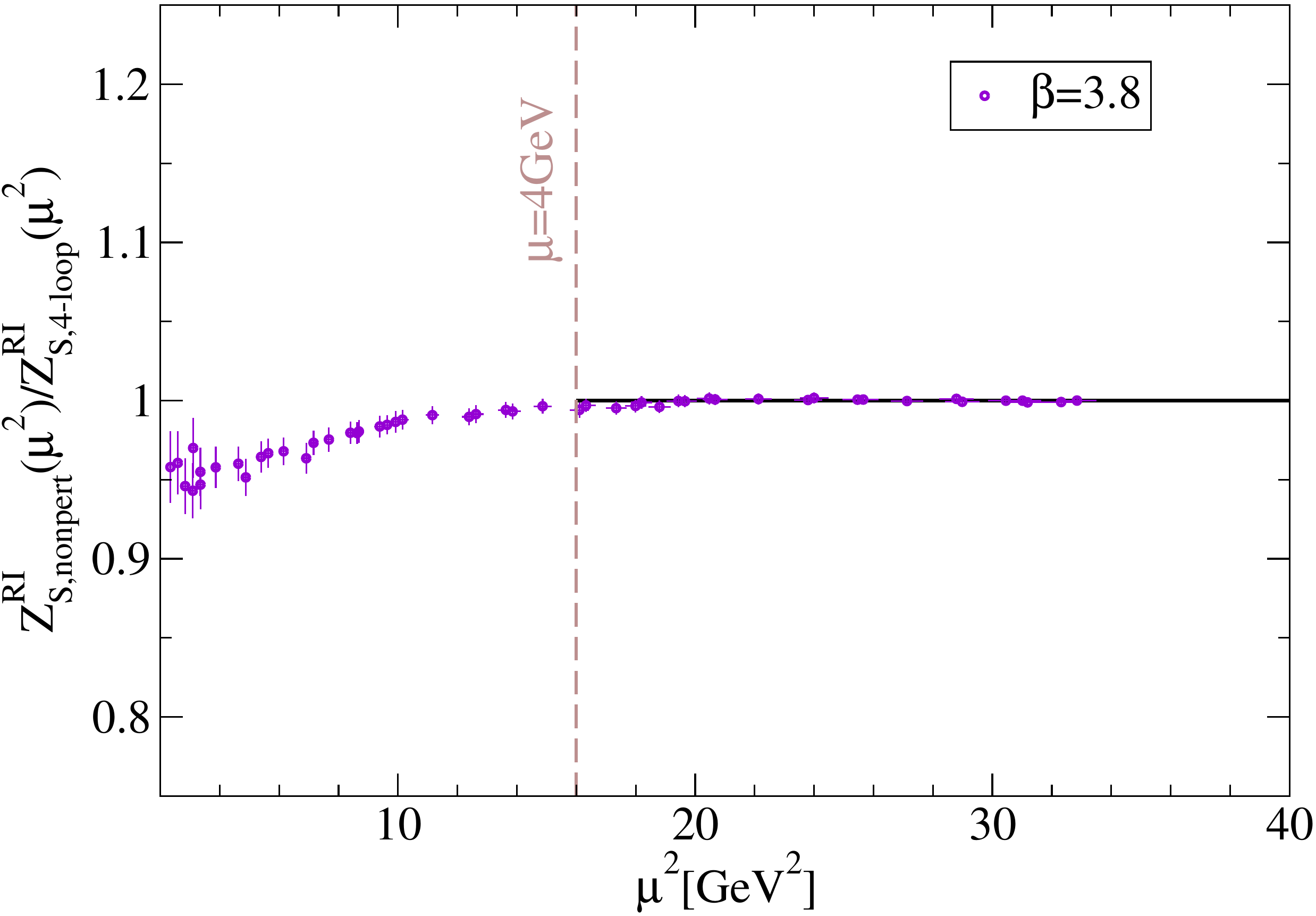}}
\caption{
  \label{fig:zsrun} Nonperturbative over 4-loop perturbative running of
  the inverse of the mass renormalization constant $Z_S$ in the RI-MOM
  scheme. Above 4~GeV agreement is reached within the statistical
  precision of our data.}
\end{figure*}

As a last step, we convert the individual quark masses from the
nonperturbative RI-MOM scheme into the $\overline\mathrm{MS}$
scheme. In both schemes, the running of the quark mass is known to
4-loop order.\cite{Vermaseren:1997fq,Chetyrkin:1999pq} As demonstrated
in fig.~\ref{fig:zsrun}, the nonperturbative running is well described
by 4-loop perturbation theory above $\mu=4$~GeV. We therefore compute
our quark masses in the RI-MOM scheme at $\mu=4$~GeV and further
convert these numbers into the $\overline\mathrm{MS}$ scheme using the
results of Chetyrkin and Retey.\cite{Chetyrkin:1999pq} Our final numbers are
\be
\begin{array}{ll}
m_u^\mathrm{\overline{MS}}(2~\mathrm{GeV})=2.17(4)(10)~\mathrm{MeV} &
m_s^\mathrm{\overline{MS}}(2~\mathrm{GeV})=95.5(1.1)(1.5)~\mathrm{MeV} \\
 m_d^\mathrm{\overline{MS}}(2~\mathrm{GeV})=4.79(7)(12)~\mathrm{MeV} 
&m_{ud}^\mathrm{\overline{MS}}(2~\mathrm{GeV})=3.469(47)(48)~\mathrm{MeV} 
\end{array}
\ee

\section{Neutral kaon mixing}

Neutral kaon mixing is responsible for indirect CP violation in
$K\rightarrow 2\pi$ decays. It is phenomenologically described by the
parameter $\epsilon$, which contains the hadronic matrix element of
the standard model $\Delta S=2$ operator $O^{\Delta
  S=2}=(\bar{s}\gamma^\mu_Ld)(\bar{s}\gamma_{\mu L}d)$ that is usually
parameterized as
\be 
B_K=\frac{\langle\bar{K}^0|O^{\Delta
    S=2}|K^0\rangle}{\frac{8}{3}\langle\bar{K}^0|A^\mu|0\rangle\langle
  0|A_\mu|K^0\rangle}
\ee
A precise determination of $B_K$ together with an experimental
measurement of $\epsilon$ thus constitutes a precision test of the
standard model in the kaon system which is particularly relevant for
constraining various standard model
extensions.\cite{Csaki:2008zd,Bauer:2011ah,Kadota:2011cr}

\begin{figure*}
\centerline{\includegraphics*[width=12cm]{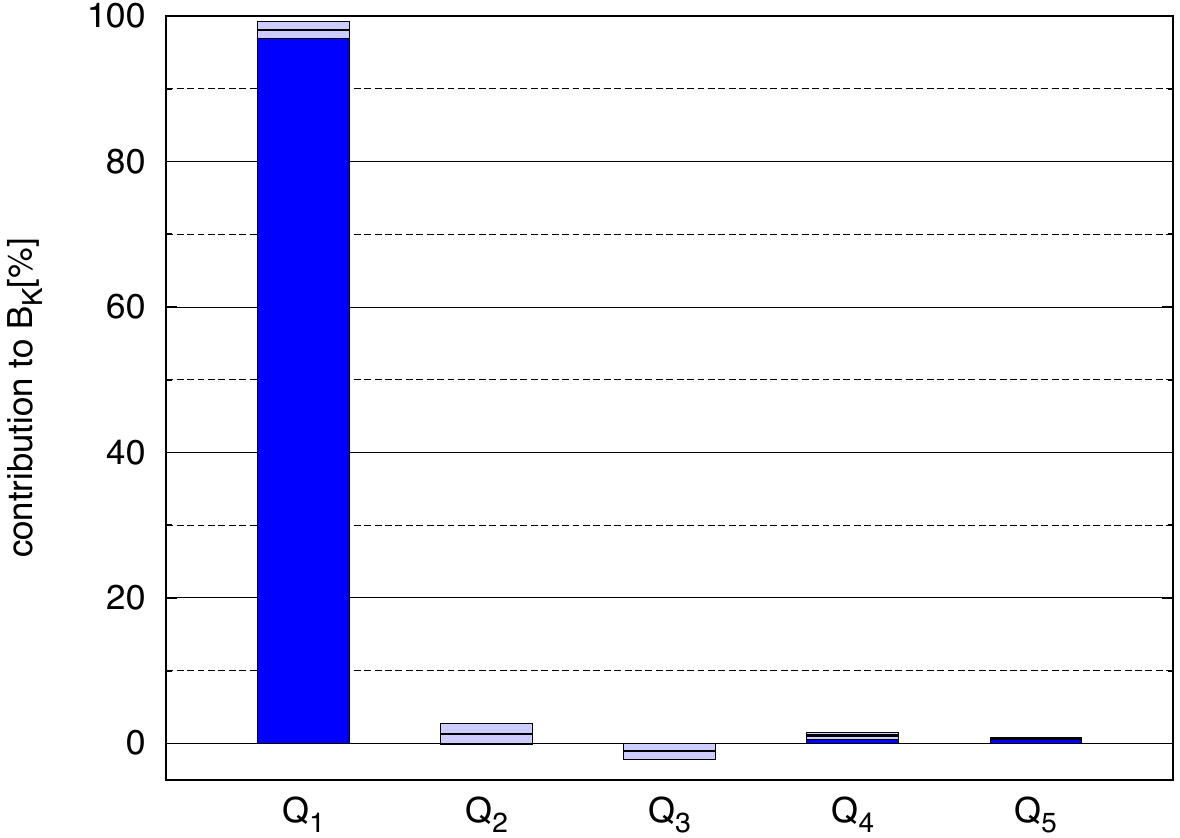}}
\caption{
  \label{fig:bkmix}Contribution of the various operators towards the
  final value of $B_K$ for the ensemble with the lightest pion
  mass. The contribution from the standard model operator $O_1$ is
  clearly dominant.}
\end{figure*}

We have performed a lattice determination of $B_K$ using the same
setup as for our quark mass determination.\cite{Durr:2011ap} One
particular point to note is that our lattice discretized fermion
action does only exhibit approximate chiral symmetry that gets fully
restored in the continuum limit only. Consequently, mixing of the
standard model operator, which has the structure $O_1=(V-A)(V-A)$, with
other dimension-6 operators that is forbidden in the continuum is
allowed at finite lattice spacing. These other operators are
$O_2=VV-AA$, $O_{3/4}=SS\mp PP$ and $O_5=TT$. As the standard model
operator is chirally suppressed, these mixings can in principle be
very large. Due to the good approximate chiral symmetry of our action,\cite{Capitani:2006ni}
the mixing contributions to $B_K$ are actually tiny as displayed in
fig.~\ref{fig:bkmix}.

\begin{figure*}
\centerline{\includegraphics*[width=8cm]{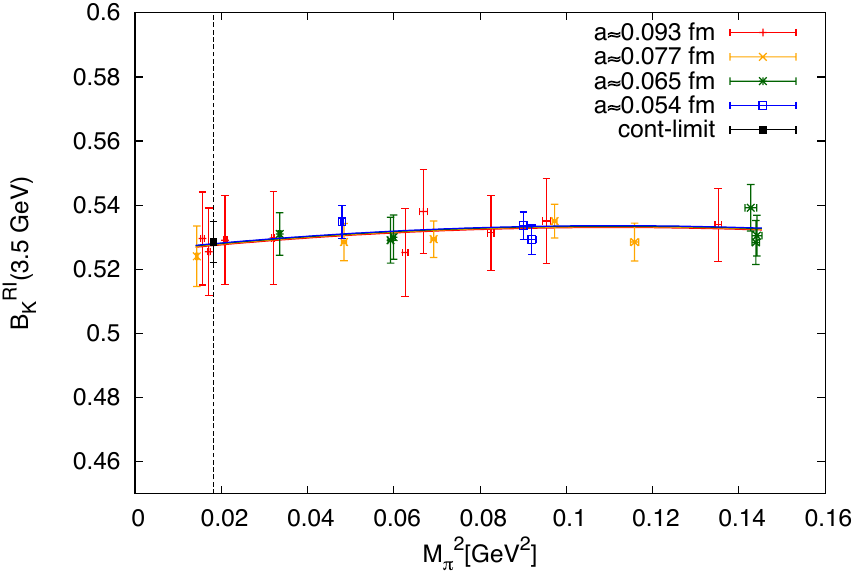}\includegraphics*[width=8cm]{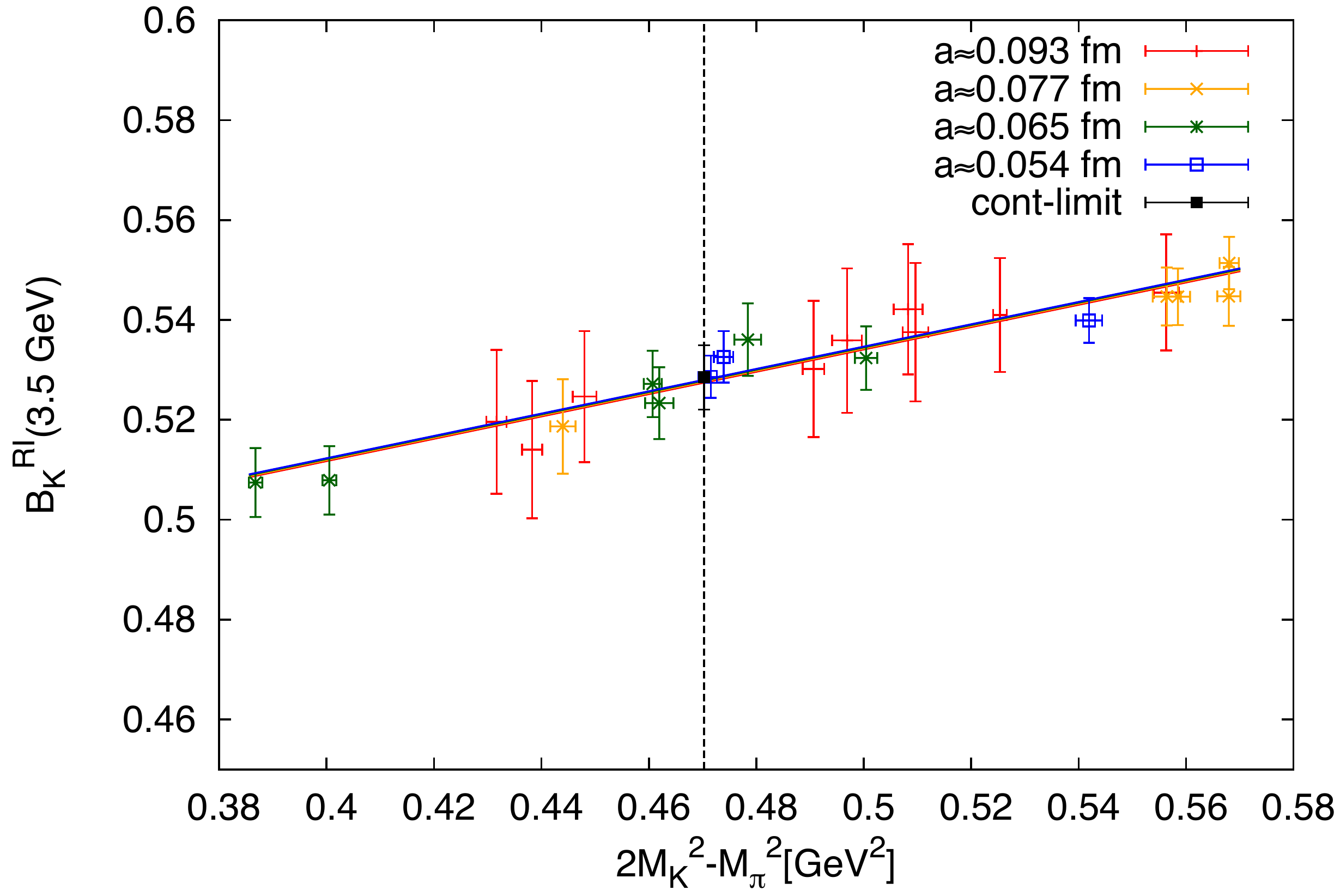}}
\caption{
  \label{fig:bkphpt}Interpolation of the renormalized lattice results
  for $B_K$ to the physical pion and kaon masses. Note that the
  interpolation curves from different lattice spacings are almost on
  top of each other.}
\end{figure*}

\begin{figure*}
\centerline{\includegraphics*[width=14cm]{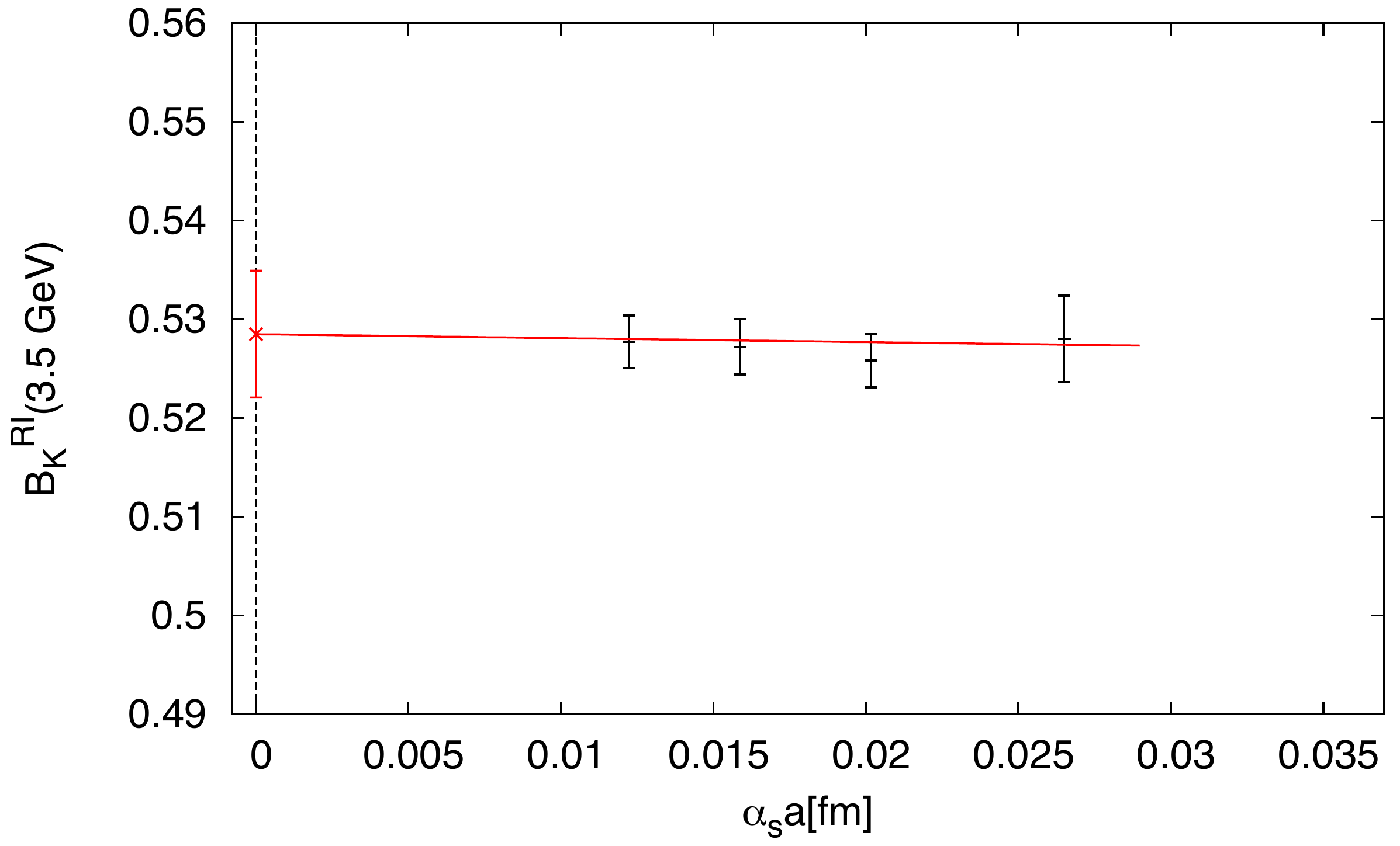}}
\caption{
  \label{fig:bkcont}Continuum extrapolation of the renormalized lattice results
  for $B_K$ at the physical pion and kaon masses.}
\end{figure*}

We measure $B_K$ on ensembles at 4 different lattice spacings and a
variety of pion and kaon masses. Renormalization is again performed
nonperturbatively in the RI-MOM scheme.\cite{Donini:1999sf} For each
lattice spacing, we interpolate the renormalized $B_K$ to physical
pion and kaon masses using various interpolators (see
fig.~\ref{fig:bkphpt}) and the resulting physical value is
extrapolated to the continuum (see fig.~\ref{fig:bkcont}).\footnote{In
fact, both the interpolation to the physical point and the continuum
extrapolation are technically performed in one combined, global fit.} In addition
to the finite volume corrections on pion masses, we also
apply finite volume corrections to $B_K$.\cite{Becirevic:2003wk}

Both interpolation to physical pion and kaon masses as well as the
continuum extrapolation turn out to be very mild. In addition, the
effect of finite volume corrections is even smaller than it was on
quark masses. Consequently, the systematic error of our result is less
than half the statistical error and we obtain
\be
\label{eq:bkri}
B_K^{\mathrm{RI-MOM}}(3.5GeV)=0.5308(56)(23)
\ee
as our final, fully nonperturbative result.

For further conversion of (\ref{eq:bkri}) into other schemes, we use
results for the 2-loop running.\cite{Ciuchini:1997bw,Buras:2000if}
Adding a conservative perturbative conversion uncertainty of $1\%$, we
obtain
\bea
B_K^{\mathrm{\overline{MS}-NDR}}(2GeV)&=&0.5644(59)_\mathrm{stat}(25)_\mathrm{sys}(56)_\mathrm{PT}\\
\hat{B}_K&=&0.7727(81)_\mathrm{stat}(34)_\mathrm{sys}(77)_\mathrm{PT}
\eea
The latter is compatible with the prediction
$\hat{B}_K=0.83{+0.21\atop -0.15}$ from a global CKM
fit.\cite{Charles:2004jd}

\section*{Acknowledgments}
I would like to thank the organizers of the $47^\mathrm{th}$ ``Recontres
de Moriond'' for creating a very pleasant and stimulating conference
atmosphere. This work has been supported by the DFG grant SFB-TR
55.

\end{document}